\documentclass[fleqn,usenatbib]{mnras}

\usepackage[T1]{fontenc}

\DeclareRobustCommand{\VAN}[3]{#2}
\let\VANthebibliography\thebibliography
\def\thebibliography{\DeclareRobustCommand{\VAN}[3]{##3}\VANthebibliography}
\usepackage{newtxtext,newtxmath}

\usepackage{graphicx}	
\usepackage{float}
\usepackage{amsmath}	
\usepackage{xcolor}
\usepackage{booktabs}

\title[Discovering Ca II with Neural Network]{Discovering Ca II Absorption Lines With a Neural Network}

\author[Xia et al.]{
Iona Xia$^{1,2}$\thanks{E-mail:ionaxia2013@gmail.com},
Jian Ge$^{3}$\thanks{E-mail:jge@shao.ac.cn},
Kevin Willis$^{1}$
and Yinan Zhao$^{4}$
\\
$^{1}$ Science Talent Training Center, Gainesville, FL 32606, USA \\
$^{2}$ Monta Vista High School, Cupertino, CA 95014, USA \\
$^{3}$ Division of Science and Technology for Optical Astronomy, Shanghai Astronomical Observatory, Chinese Academy of Sciences,
Shanghai 200030, China\\
$^{4}$ Department of Astronomy of the University of Geneva, 51 chemin de Pegasi, Versoix 1290, Switzerland
}

\date{Accepted 2022 October 5. Received 2022 October 2; in original form 2022 August 22}

\pubyear{2022}

\begin{document}
\label{firstpage}
\pagerange{\pageref{firstpage}--\pageref{lastpage}}
\maketitle

\begin{abstract}
Quasar absorption line analysis is critical for studying gas and dust components and their physical and chemical properties as well as the evolution and formation of galaxies in the early universe. Ca II absorbers, which are one of the dustiest absorbers and are located at lower redshifts than most other absorbers, are especially valuable when studying physical processes and conditions in recent galaxies. However, the number of known quasar Ca II absorbers is relatively low due to the difficulty of detecting them with traditional methods. In this work, we developed an accurate and quick approach to search for Ca II absorption lines using deep learning. In our deep learning model, a convolutional neural network, tuned using simulated data, is used for the classification task. The simulated training data are generated by inserting artificial Ca II absorption lines into original quasar spectra from the Sloan Digital Sky Survey (SDSS) whilst an existing Ca II catalog is adopted as the test set. The resulting model achieves an accuracy of 96\% on the real data in the test set. Our solution runs thousands of times faster than traditional methods, taking a fraction of a second to analyze thousands of quasars while traditional methods may take days to weeks. The trained neural network is applied to quasar spectra from SDSS’s DR7 and DR12 and discovered 399 new quasar Ca II absorbers. In addition, we confirmed 409 known quasar Ca II absorbers identified previously by other research groups through traditional methods.

\end{abstract}




\section{Introduction}

Quasar absorption lines are critical to study gas and dust components in galaxies and galaxy evolution as these lines can be analyzed to investigate the physical properties and kinematics of interstellar medium in their associated galaxies at different redshifts \citep{Sardane2014}. Most previous studies involving quasar absorbers concentrate on absorption lines at higher redshifts such as Magnesium II (Mg II) absorbers and damped Ly$\alpha$ absorbers (DLAs). The rare Calcium II (Ca II) absorbers are the main type of absorbers that allow the study of environments with absorption redshifts $z_{abs}$ of $<$0.4 (i.e., the most recent 4.3 billion years), as the Ca II H\&K doublet at wavelengths of \(\lambda\)\(\lambda\) 3934, \(\lambda\)\(\lambda\) 3969 can be detected in quasar optical spectra to trace galaxies at a redshift range of 0 < \(z\) < 1.4, which covers galaxies from about 8.9 billion years ago to the present epoch. Previous studies show that most Ca II absorbers lie in areas that are dense and dusty, full of metal and molecular hydrogen, the perfect spots for star formation \citep{Wild2005, Wild2006, Wild2007, Zych2007, Nestor2008, Zych2009, Sardane2014}.

However, Ca II absorption lines are uncommon because calcium is a highly refractory element, causing them to be depleted in the interstellar medium (ISM) \citep{Savage1996, Wild2005, Wild2006, Sardane2014}. Ca\textsuperscript{+} is also not the most dominant ionization state of calcium as the ionization potential of Ca II is only 11.871 electron volts (eV) while the ionization potential of H is 13.598 eV \citep{Moore1970}. With the ionization potential of Ca II being less than that of H, it is not able to be shielded by hydrogen in the ISM. With both of these factors combined, Ca II absorption lines are extremely rare. In fact, compared to the more common Mg II absorption lines, only about 3\% of quasar spectra containing Mg II absorption lines contain Ca II absorption lines \citep{Sardane2014}. Therefore, there are currently a relatively small number of quasar spectra with Ca II absorption lines discovered, with the largest catalog being from \citet{Sardane2014} with 435 quasar Ca II absorbers. This situation limits the number of statistical tests that can be accurately performed, causing a challenge to confirm previous theoretical models on physical processes and environments associated with these absorbers and understand their properties and evolution with redshifts.

Currently, a few groups have formed theoretical models on how Ca II absorption lines form in ISM associated with distant galaxies. For instance, \citet{Sardane2014} found that quasar spectra with Ca II absorption lines may be composed of two distinct populations, which arise when the distribution of equivalent width (EW) of the \(\lambda\)3934 line requires a double exponential function to produce a satisfactory fit of the data \citep{Sardane2014}. They inferred that the split of the two populations is when the ratio of EW of the \(\lambda\)2796 (Mg II absorption line) and the EW of the \(\lambda\)3934 (Ca II absorption line) equals 1.8, and at the EW of the Ca II absorption line at \(\lambda\)3934 = 0.7\r{A}, with absorption lines that have a \(\lambda\)3934 EW below 0.7\r{A} considered as weak absorbers, and absorption lines that have a \(\lambda\)3934 EW above 0.7\r{A} considered as strong absorbers. The weak population can be identified with a halo-like gas, while the strong population can be identified with a mix of halo and disk-like gas. Investigations on chemical and dust depletion properties of Ca II absorption lines subsamples done by \citet{Sardane2015} and \citet{Fang2022} have results supporting that stronger absorbers are likely to be associated with low impact parameter and disc-like environments compared to the weaker absorbers’ association with larger impact parameter environments typical of galactic halos. Strong absorption lines are also found to be six times more reddened than weak absorption lines. They also noted an imbalance between the two populations with there being a much larger amount of weak absorbers.

Another theoretical model that was postulated by \citet{Wild2005} speculates that Ca II absorbers are an unusual subclass of DLA systems, as Ca II absorbers have similar strengths of neutral hydrogen column densities of various lines of sight through the Milky Way galaxy as DLAs. In a study by \citet{Zych2007}, they found that 18 of the 19 Ca II quasar absorbers studied have <0.3 dex variation in [Cr/Zn], much like the general DLA population, providing more evidence for this model. However, data for verifying these models are relatively small and uncertainties are large. Thus, more data is required to provide strong constrains on these models and understand their characteristics. 

There is also a possible link to molecular clouds and star forming galaxies (SFGs) which is suggested by the detection of H\textsubscript{2} absorption in a few Ca II absorbers. \citet{Sardane2015} suggested that this might be a common phenomenon and linked this with their depletion of dust grains and reddening. 
\begin{figure*}
 \includegraphics[width = \textwidth]{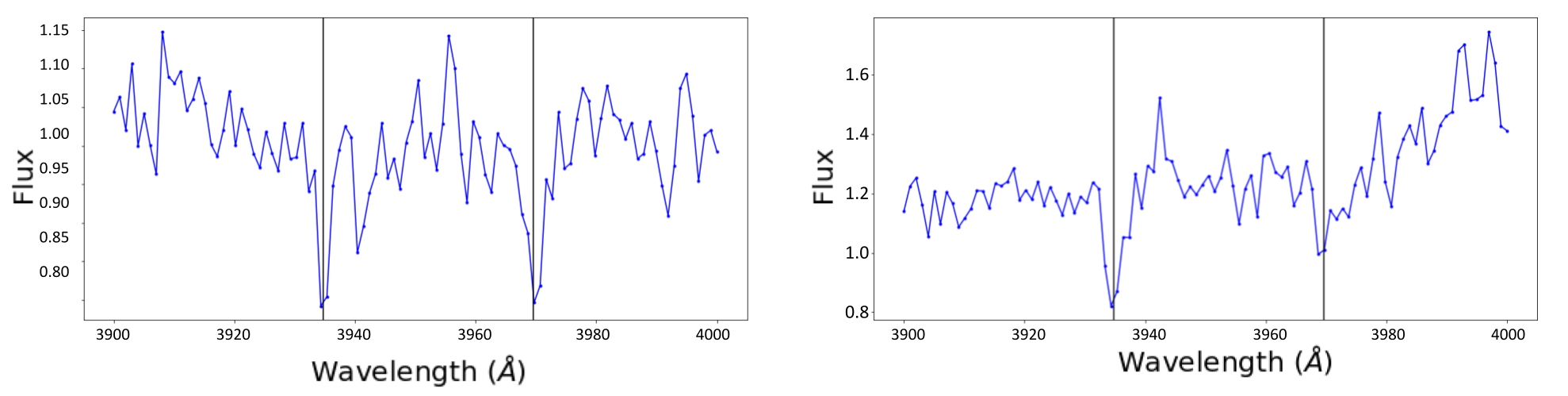}
 \caption{Examples of positive quasar spectra samples with Ca II absorption lines (from \citet{Sardane2014}) in the test set. The \(\lambda\)3934 and \(\lambda\)3969 lines are indicated by the black lines.}
 \label{fig:figtest}
\end{figure*}

Nevertheless, despite the necessity to discover more quasars with Ca II absorption lines, the traditional method of detecting them takes an extremely long time, taking usually many days to many weeks to detect absorption lines in large datasets such as the SDSS releases \citep[e.g.][]{Fang2022}. It also requires considerable human interference, making it quite inconvenient and possibly inconsistent results. The basic search procedure for Ca II absorbers is actually rather simple. As summarized in \citet{Sardane2014}, which uses methods described in \citet{Nestor2005, Rimoldini2007, Quider2011}, there is first an automated process in which the quasar spectrum is normalized and fitted using a combination of cubic splines and Gaussian functions. In order to provide a good fit, it takes significant computational time. This is not ideal as there are tens of thousands of quasar spectra that need to be processed. A sliding line-finding window is then applied on the quasar spectrum to search for the possible Ca II doublet. If it is found, and it has a sufficient signal-to-noise ratio (SNR), it is flagged as a candidate set of Ca II absorption lines. Next, the candidates are manually checked to determine if they are real Ca II absorption lines. These manual checking criteria differ between different groups, with most earlier research work seeking stronger Ca II absorption lines. For example, \citet{Zych2007} required an EW > 0.2 \r{A} and a minimum significance level of 4\(\sigma\) for the Ca II \(\lambda\)3934 line and a minimum significance level of 1\(\sigma\) for the \(\lambda\)3969 line. \citet{Sardane2014}, meanwhile, had a harsher criteria with a 5\(\sigma\) minimum level of significance for the \(\lambda\)3934 line, and a 2.5\(\sigma\) minimum level of significance for the \(\lambda\)3969 line, and also sought for a 2:1 line ratio between the \(\lambda\)3934 line and the \(\lambda\)3969 line. This step takes the largest amount of time and human interference, as often there are thousands of absorption lines that need to be double-checked one at a time. With all the time and human interference that is required to search for Ca II absorption lines, it would be extremely helpful to use a faster and more accurate method. 

Because of this, we decided to develop a deep neural network approach in order to create a novel, fast, and accurate method. A neural network is a set of algorithms that are designed to recognize patterns in a series of data in a process that mimics the human brain and is one kind of artificial intelligence. It is able to solve problems that are otherwise impossible or difficult to deal with as it can act fast like a computer and accurately like the human brain. Neural networks consist of neurons that are connected through the input layer, hidden layers, and the output layer. The two most popular neural networks that are used for data classification are the convolutional neural network (CNN) and the recurrent neural network (RNN). The CNN is adopted as it contains classification capabilities that are necessary to identify Ca II absorption lines as opposed to RNNs which are more adapted for time series data analysis \citep{Zhao2019}.

In the astronomy field, neural networks have been used in a variety of ways from 1D spectra analysis e.g., \citep[e.g.][]{Hala2014, Graff2016, Parks2018, Hampton2017} to 2D image pattern recognition in star galaxy classification \citep[e.g.][]{Bertin1996, Kim2017}. \citet{Zhao2019} developed a deep neural network for identifying Mg II \(\lambda\)\(\lambda\) 2796, 2803 \r{A} doublets in the SDSS DR12 data set and were able to achieve about a 94 percent accuracy while analyzing 50,000 quasar spectra in around 9 seconds for identifying Mg II absorber candidates.

In this paper, we aim to modify the deep neural network developed by \citet{Zhao2019} for identifying Ca II absorbers in the SDSS DR 7 \citep{Abazajian2005} and DR 12 \citep{Alam2015} data sets. As Ca II absorbers are weaker than Mg II absorbers, it is a more difficult and challenging task to identify them. In section 2, the data sets used for the training and testing purposes in this paper are discussed, including the artificial spectra generated, as well as data preprocessing for maximum accuracy and efficiency. In section 3, the steps to build our neural network are introduced and the developed model is described. Experimental results are summarized in section 4, followed by a discussion, concluding remarks, and plans for future improvements in section 5.

\section{DATA SETS AND DATA PREPROCESSING}

Because of the lack of quasar spectra with Ca II absorption lines discovered and the magnitude of data necessary to have an effective and accurate training set for the neural network model, there is a need for an artificial training data set using existing quasar data. Our artificial spectral data are generated based on SDSS DR12 \citep{Alam2015}. Nevertheless, an existing catalog of quasar Ca II absorbers from \citet{Sardane2014} is used as our test set since the different parameters and outliers (such as imperfections in real data) may be present in real data and may not be covered in the artificial spectra. This would help identify potential loose ends in our neural network design. Because of this, to make our model as practical as possible, it is beneficial to use a real data set as the test data set in order to not only test our model but also to evaluate our training data set. It is also important to note that all negative samples in our test set and all datasets used to search for Ca II absorption lines include Mg II absorption lines detected before, as our approach requires the use of \(z_{abs}\) to select the spectral region covering both Ca II absorption lines. All quasars Ca II absorption lines have been found to also have Mg II absorption lines with a \(z_{abs}\) > 0.4 and a good SNR at the Mg II absorption line location.

\begin{figure}
 \includegraphics[width = \linewidth]{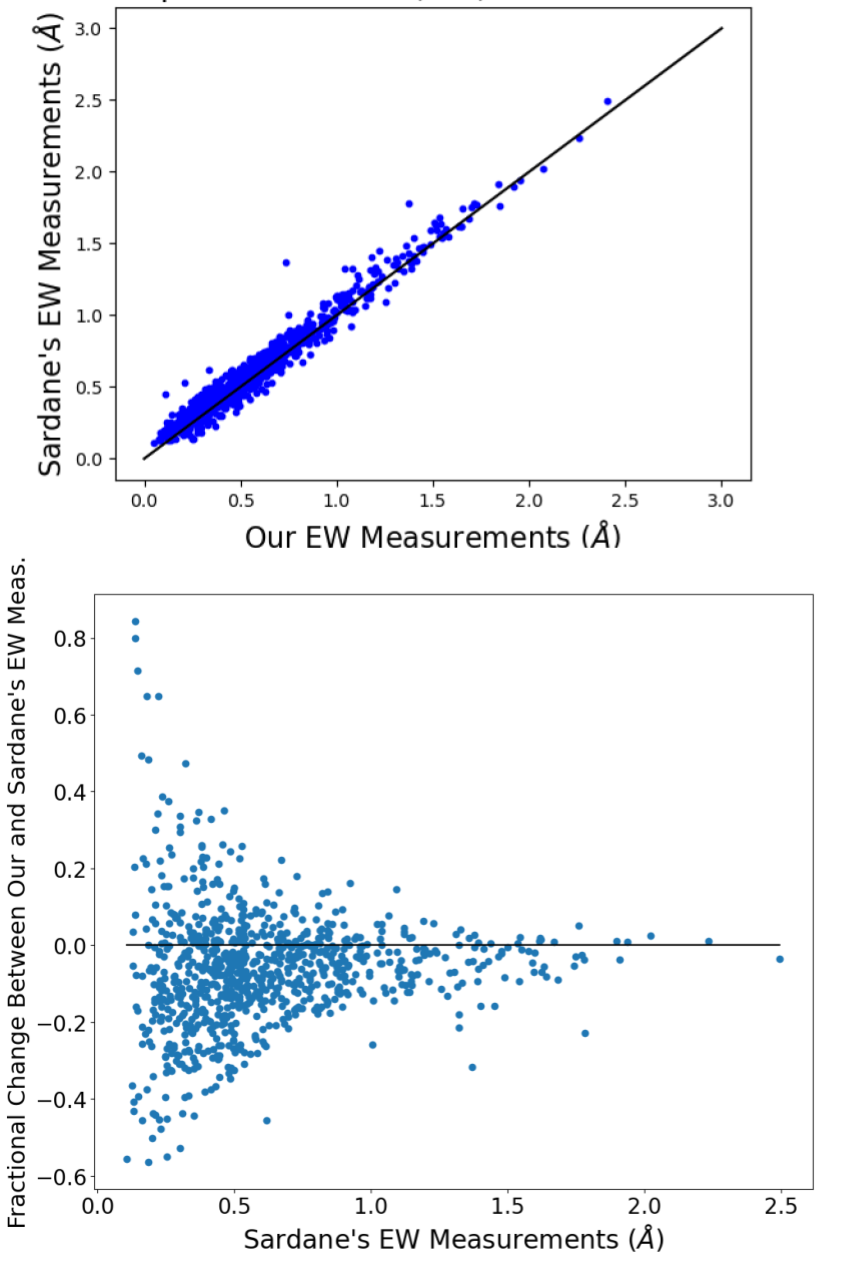}
 \caption{A comparison between our measured equivalent widths and \citet{Sardane2014}'s measurements for its catalog. While there were a couple of outliers, most of the measurements were extremely similar (being very close to the black line x=y). We find that some of the outliers may be false detections as described in 4.1. Top: A comparison with our measurements on the y-axis and \citet{Sardane2014}'s measurements on the x-axis. Bottom: A comparison between the fractional difference of our measurements and \citet{Sardane2014}'s measurements with \citet{Sardane2014}'s measurements. There was a greater fractional difference for absorption lines with a lower equivalent width due to uncertainties in continuum fitting for weak absorption lines. We also found that our measurements were slightly lower overall which may be due to different approaches in spectral continuum fitting adopted by us and \citet{Sardane2014}.}
 \label{fig:figmeas}
\end{figure}
\begin{figure*}
 \centering
 \includegraphics[width = \textwidth]{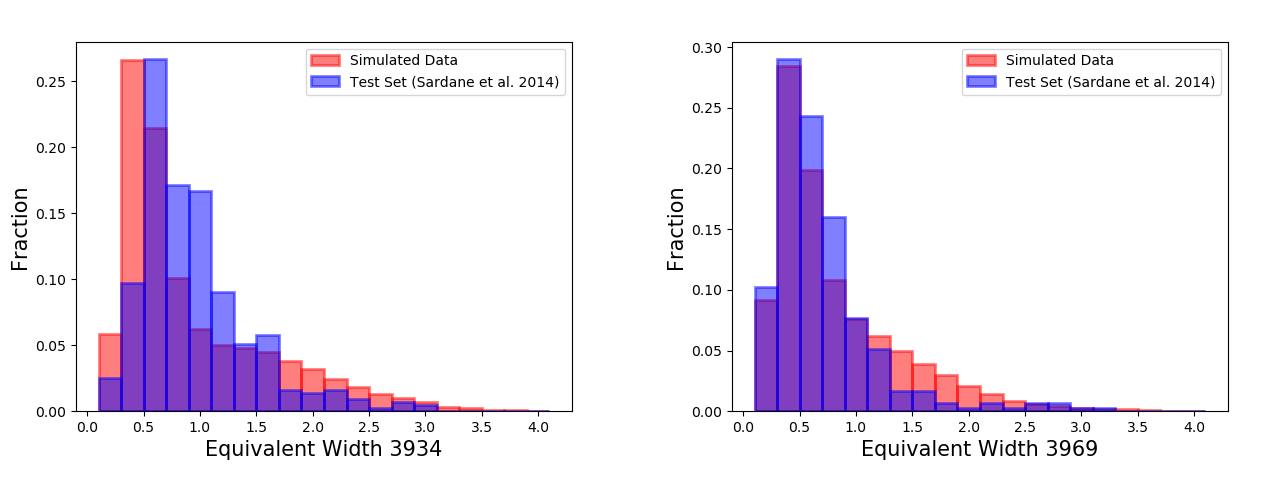}
 \caption{The distribution of the equivalent widths (EWs) for the \citet{Sardane2014}'s catalog which we used as a guiding set compared to the distribution of our simulated data set, with the red bars being the simulated data and the blue bars being the real data. Left: The EW distributions of the \(\lambda\)3934 line. The distributions of both datasets are similar and have the same trend. However, our simulated data has a greater amount of lower EWs for the \(\lambda\)3934 line so it can be more sensitive to samples with low EWs. Right: The EW distributions of the \(\lambda\)3969 line. The distributions of the two datasets are very similar.}
 \label{fig:figew}
\end{figure*}
\subsection{Real Ca II Absorption Lines and Test Data Set}

 431 SDSS quasar spectra in the quasar Ca II absorber's catalog of \citet{Sardane2014} are used as the positive samples, making half of our test data set. Only 4 samples in this catalog were missed because we were unable to locate the mjd, fiber, and plate for them. This catalog has quasar Ca II absorbers with 0.0277 < \(z_{abs}\) < 1.3428, with a mean \(z_{abs}\) of 0.5747. Their catalog was chosen with a 5\(\sigma\) minimum level of significance for the \(\lambda\)3934 line, and a 2.5\(\sigma\) minimum level of significance for the \(\lambda\)3969 line. They also mentioned that their dataset should be very complete with no more than 10 absorption lines missed  \citep{Sardane2014}. These Ca II absorbers were found from quasar spectra from SDSS DR 9 \citep{Ahn2012} and DR 7 \citep{Abazajian2005}.
 
 In addition to that, we randomly chose 431 other quasar spectra that do not contain Ca II absorption lines as the negative half of our test data set. A 1:1 ratio between the sizes of the positive and negative data sets was adopted as studies have shown that having equal numbers of positive and negative samples is the most optimal way to train and test neural networks \citep{Masko2015TheIO}. Some examples of positive samples in the test set are shown in Figure \ref{fig:figtest}.
\begin{figure}
 \includegraphics[width = 0.47\textwidth]{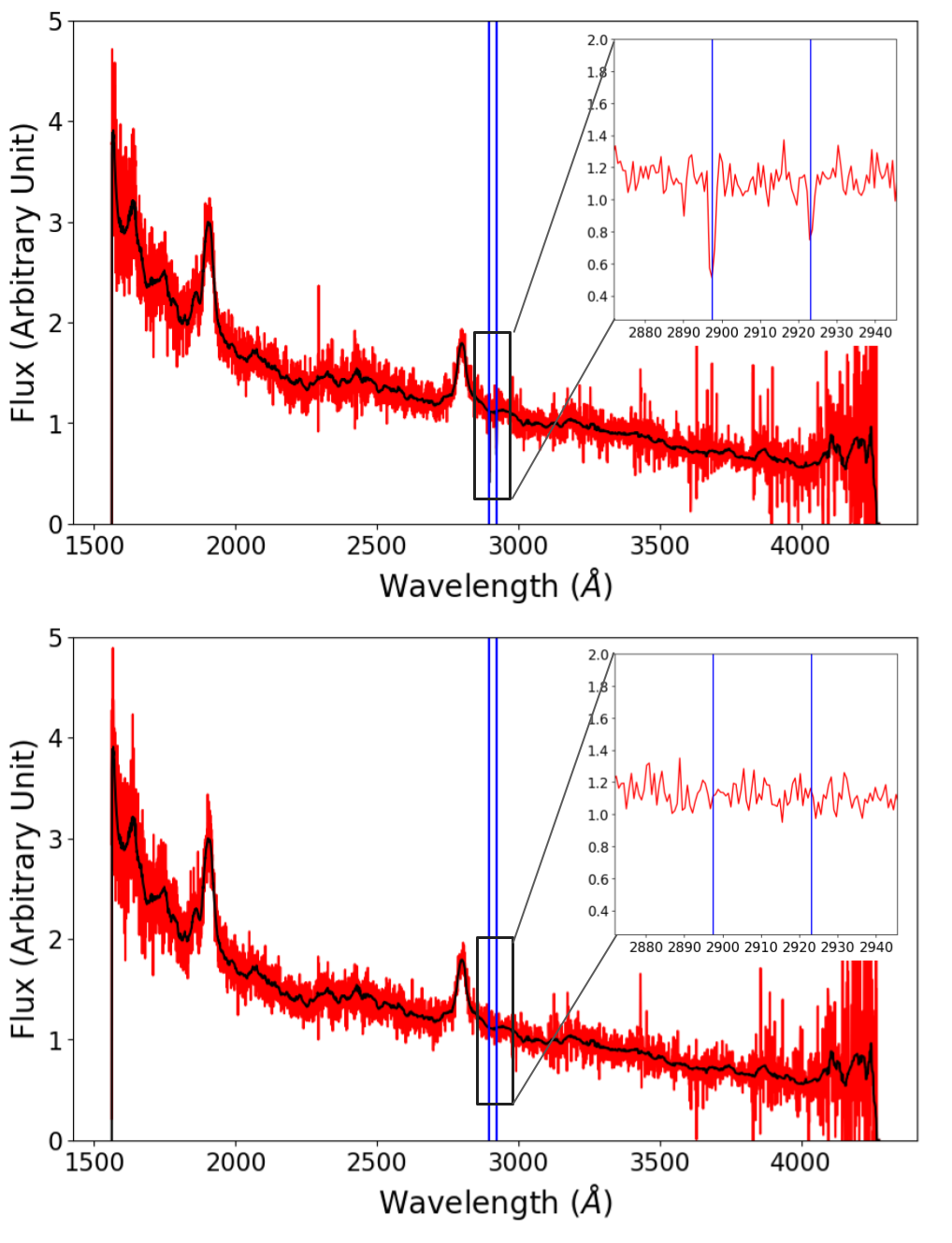}
 \caption{Examples of the artificial spectra generated. For each example, the whole spectrum is shown in the bigger box with the red line being the flux and the black line being the continuum fit of the spectrum. A closeup of the normalized areas around the Ca II absorption lines is shown in the smaller box on the top right of the plot. The blue lines mark where Ca II absorption lines are inserted. Top: A positive artificial spectrum. Bottom: A negative artificial spectrum made using the same original spectrum as the one shown above.}
 \label{fig:figart}
\end{figure}
\subsection{Artificial Ca II Absorbers and Training Data Set}
\begin{figure*}
 \includegraphics[width=0.8\textwidth]{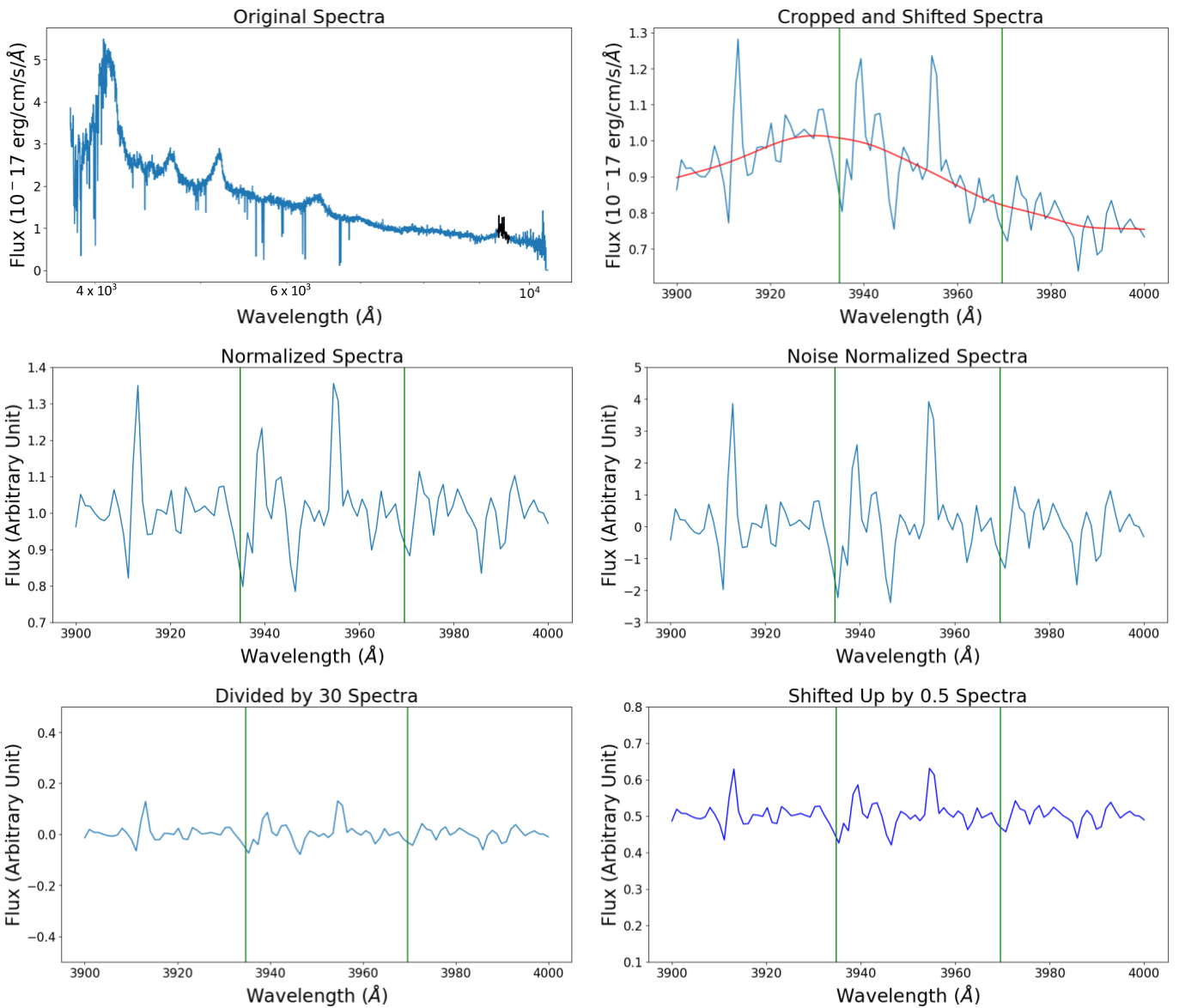}
 \caption{A demonstration of the procedure used for preprocessing. Top Left: The original SDSS quasar spectrum with Ca II absorption lines with the wavelength in log scale. The part of the spectrum (3900 \r{A} to 4000 \r{A}) that is actually preprocessed is shown in black color. Top Right: The spectrum after the original spectrum is shifted to the rest frame and cropped, with its continuum fit shown as a red line. Middle Left: The cropped spectrum is normalized. Middle Right: The normalized spectrum is subtracted by 1, then divided by the standard deviation of the flux in the wavelength window except for 3930-3938 \r{A} and 3965-3973 \r{A} to have a noise normalized spectrum. Bottom Left: The noise normalized spectrum is divided by 30. Bottom Right: 0.5 is added to the processed spectrum to have the desired flux range from 0.0 - 1.0, which has an optimal sensitivity for feeding our trained NN.}
 \label{fig:figpp}
\end{figure*}
Since it is essential to produce an abundance of training data to train an accurate neural network, artificial quasar spectra were generated. In order for the artificial data to be representative of real quasar spectra with Ca II absorption lines, artificial Ca II absorption lines need to follow parameter distributions of real Ca II absorption lines. 
Therefore, measurements of key parameters describing Ca II absorption lines, such as the EW, the full width at half maximum (FWHM), and \(z_{abs}\) in real Ca II absorber data were conducted. The measurements are used for generating artificial Ca II absorption lines.

The EW values from Sardane's catalog were remeasured by our program to be consistent with our measurements for new Ca II absorbers, even though \citet{Sardane2014} listed the EWs of their discovered absorbers in their paper (Figure \ref{fig:figmeas}). When comparing our measurements with theirs, we found that our equivalent width measurements were slightly lower overall by about 6\%. This could be due to our difference in  continuum fitting. Figure 5 shows an example of our continuum fitting which filters out above 2\(\sigma\) outliers of spectral data points before fitting the continuum. Sometimes, it is necessary to have multiple iterations in fitting before converging to a final continuum solution. Since a typical measurement error for most weak absorption lines is on the order of 20-25\%, this 6\% systematic small offset is a factor of at least three times smaller than these measurement errors. Therefore, our EW measurements are consistent with the measurements by Sardane et al.

In addition, the FWHM values of their absorbers were also measured by first cropping the spectrum to where the absorption line is, by shifting the spectrum to the rest frame and then cropping it to 3900 \r{A} to 4000 \r{A}. Then, the remaining segment was normalized by fitting it to a continuum using a polynomial function. Our absorption fitting program is based on a Python Voigt fitting package called VoigtFit (\citet{Krogager2018}) which can measure both the FWHM and the EW of absorption lines. The FWHMs follow a Gaussian distribution of around \(\mu\) = 2.1 \r{A} and \(\sigma\) = 0.9 \r{A}. The EW distribution has a range between 0.16 and 2.57 \r{A} for the \(\lambda\)3934 line with an average equivalent width of 0.76 \r{A}, while the \(\lambda\)3969 line has an EW distribution of between 0.11 and 1.60 \r{A} with an average of 0.48 \r{A}. Their \(z_{abs}\) values are between 0.03 and 1.34.

Our artificial data set follows a similar distribution from 0.05 to 4.00 \r{A} for the EW to account for possible stronger absorption lines, but with a greater amount having a lower EW to match the distribution of the real dataset. The comparisons of the distributions are shown in Figure \ref{fig:figew}. As can be seen, the distributions of the two datasets are similar and have the same trend for both \(\lambda\)3934 and \(\lambda\)3969 lines, indicating that a randomly selected sample from the artificial data would roughly match the distribution of \citep{Sardane2014}’s catalog. However, the simulated set contains a greater amount of lower EWs for the \(\lambda\)3934 line, so the trained neural network can be more sensitive to samples with low EWs. As for \(z_{abs}\), because we are searching quasar spectra with detected Mg II absorption lines, the redshift distribution was changed to 0.36 < \(z_{abs}\) < 1.4. In other words, we use redshifts of Mg II absorbers to search for potentially corresponding Ca II absorbers at the same redshifts, which makes the search much quicker and more accurate than a blind search, but loses the capability in searching for Ca II absorbers at redshifts lower than 0.36. We also explored the Sloan quasar spectra and noted that there is heavy contamination of high sky noises in quasar spectra with observer wavelengths greater than 8500\r{A} due to strong sky emissions lines. However, our method with normalization of data noise in preprocessing as described in section 2.3 is able to reliably identify Ca II absorbers in these noisy spectra.

The base data for our artificial data set is drawn from DR12 of SDSS \citep{Alam2015}. We randomly chose a subset of around 12,000 spectra that had the correct \(z_{qso}\) range from the DR12 SDSS catalog, making sure that none are part of our test set. Because an even larger training set is desired, six variations are created for every spectrum, each with a different redshift and EW of the inserted lines. When inserting an absorption line, the method used in \citet{Zhao2019} was adopted, which involved first performing a continuum fit for each spectrum, injecting artificial Ca II absorption lines to each continuum, and then adding noise after that. The continuum fitting to quasar spectra was conducted using the principal component analysis (PCA) method, which is a commonly used unsupervised machine learning algorithm for data de-noising and smoothing. The entire quasar sample set was divided into a number of subsets by the quasar’s emission redshift with a bin size of 0.2, and then performed the fitting to each subset and derived the Eigen spectra. 

The Ca II absorption lines are inserted at their specified locations of 3934.78 \r{A} and 3969.59 \r{A} of a quasar spectrum at a redshift in the observer frame based upon FWHM and EW distributions of real Ca II absorbers. The redshift is randomly selected in the range stated above. The quasar spectrum's SNR around the Ca II absorption line area must meet the minimum requirement of 3.0 measured with the spectrum flux and error arrays. In addition, a SNR greater than 2.0 must be met for both of the inserted \(\lambda\)3934 and \(\lambda\)3969 lines. The EW error is computed as the multiplication of the FWHM by the local noise. In total, 72,000 artificial spectra were generated with lines inserted, and another 72,000 spectra without line insertion, which we had created in the same way that we created the 72,000 positive samples by randomly selecting six different redshifts for each of the 12,000 real spectra. Examples of these \textit{positive} and \textit{negative} samples are shown in Figure \ref{fig:figart}. This dataset is randomly divided between the training and validation set with a ratio of 4:1 for neural network training.

\subsection{Data Preprocessing}

An optimized data preprocessing method was developed to improve the neural network's accuracy and efficiency. In this method, the quasar spectra are shifted to the absorber rest frame for searching for Ca II absorption lines. By doing this, the absorption lines are always at the same location on every spectrum and thus, it is easier for the neural network model to be trained and used to find them. This allows cropping the quasar spectra to a relatively small fixed wavelength range between 3900 \r{A} and 4000 \r{A} for data processing and absorption line search, which not only decreases the processing time and training time but also allows the model to be focused on the region of importance, therefore making it easier for the model to locate the absorption lines and increasing the accuracy. On the other hand, this wavelength range is still large enough so that the model is able to evaluate and consider local features in spectra to assist its determination. Therefore, in order to make this method work, the absorber's redshifts of all spectra that are analyzed by the model are required to be known so that the spectra can be shifted to the rest frame. 

In the processing, each spectrum is first normalized through continuum fitting. If absorption lines are close to or on quasar emission lines at \(z_{qso}\), a flexible smoothing spline function is used to fit and normalize the continuum. Otherwise, they are normalized with a polynomial fit. In the second step, a constant of 1 is subtracted from the normalized spectrum, then the spectrum is divided by the standard deviation of the spectrum between 3900 \r{A} and 4000 \r{A} except for 3930 to 3938 \r{A} and 3965 to 3973 \r{A} to normalize the noises in this spectral region. In the third step, the noise normalized spectrum is further divided by a constant value, $A$, then added by a constant offset, $B$, to complete the processing. We divided it by $A$, so that all the flux values are scaled to be within a small enough range, with the maximum difference between any two samples not greater than 1.0. We then added it by the constant offset, $B$, so that it would have the desired relative intensity range from 0.0 - 1.0. Many different combinations of $A$ and $B$ were tried, and values of $A=30$ and $B=0.5$ appear to make the model more sensitive for detecting Ca II absorption lines than other combinations. This process is illustrated in Figure \ref{fig:figpp}, which is especially helpful to search for Ca II absorption lines in spectral regions with high sky noises due to sky emission lines. This helps extend the Ca II absorber's searching redshifts to higher values. This prepossessing has been applied to all the 862 spectra in the test set and the 144,000 spectra in the training set. In the last step, the positive and negative samples in the data were separated with the positive samples labeled as 1, and the negative samples labeled as 0 for both the training and testing sets to complete the processing.

\section{NEURAL NETWORK MODEL STRUCTURE}

\begin{figure*}
 \includegraphics[width=\linewidth]{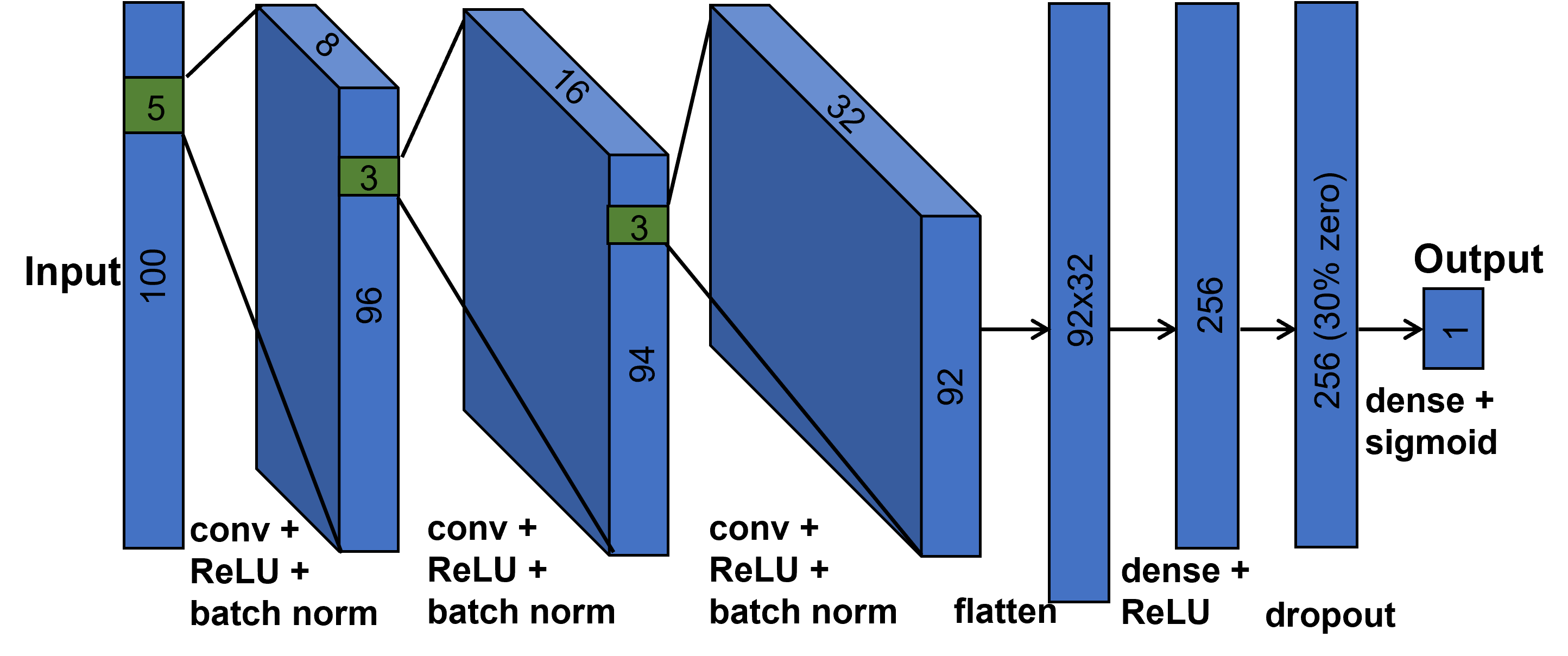}
 \caption{A diagram of the developed neural network model featuring 3 convolutional layers, 2 fully-connected layers, and a dropout layer with an input spectrum of 100 pixels.}
 \label{fig:figarc}
\end{figure*}

We chose to use a convolutional neural network (CNN) model, which is designed to search within a fixed wavelength window of each input spectrum. This significantly simplifies the task and the model, as shown in Figure \ref{fig:figarc}, allowing for fewer model layers and significantly reducing processing time.  Our model includes convolutional layers, fully connected layers, dropout layers, activation layers and normalization layers.

Convolutional layers are the main components of a CNN, with the layers applying a number of filters to the input. These filters will then create feature maps that represent certain features of the input data. In our model, the convolutional layers served to identify patterns of the spectral lines. Key characteristics of these layers include the kernel shape and quantity, which were determined through hyperparameter tuning, and further explained in section 3.1. Likewise, convolutional layer quantity was also selected in a similar fashion.

Following each convolutional layer are activation and batch normalization layers. Activation layers are used to introduce non-linearity to the model so that it is able to learn and represent the complexity in the data patterns. The batch normalization layers are applied to ensure that inputs to the next layer are scaled to the standard normal distribution, giving the model training stability and consistency. A max pooling layer was also considered following each convolutional layer, to reduce the amount of input data to the next layers. A dropout layer is often included towards the end of the model to randomly set a certain amount of input units to zero in order to prevent the model from overfitting.  Overfitting would cause the model to have a hard time predicting anything other than the training set because it is too attuned to the specific training set's detailed features, including noises. Fully connected layers are finally used at the end after extracting the features with the convolutional layers to classify features and get the final result.

\begin{figure*}
 \includegraphics[width=0.9\textwidth]{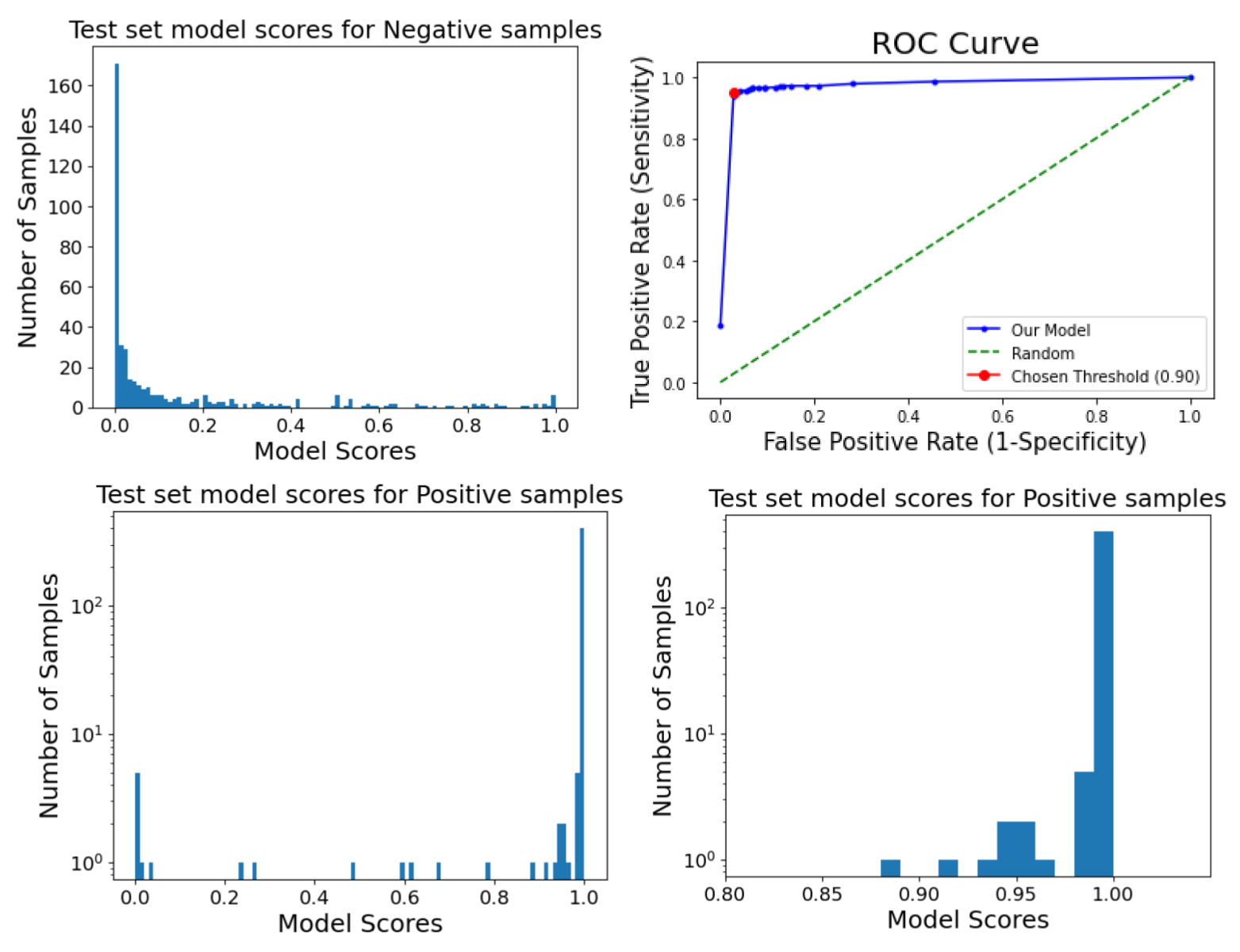}
 \caption{The output scores from the model given to negative samples and positive samples from the test set and the Receiver Operating Characteristic (ROC) curve of the model. Top left: Model scores for negative samples in the test set. Although most of the scores are below 0.5, there are still quite a few of them higher than 0.5. Top right: The ROC curve of the model shows its sensitivity and specificity at different threshold values (from 0 to 1). Bottom left: Model scores for positive samples in the test set. The vast majority of them are between 0.8 and 1.0. The number of samples are shown in log scale to better represent the distribution. Bottom Right: A closer look at the model scores within [0.8, 1.0] for positive samples, it can be seen that most of them lie greater than 0.90, which became the selected threshold for the model. The number of samples are shown in log scale.}
 \label{fig:figthreshold}
\end{figure*}
\begin{table*}
 \caption{Examples of model hyperparameters tested and selected, where the selected choice is marked by an "X".}
 \label{tab:hyperparameters}
 \centering
 \begin{tabular}{|cc|cc|cc|}
 \hline
 \textbf{Max Pooling} & &\textbf{Conv. Layers}& &\textbf{Output Activation}& \\
 \hline 
 1&X&1&&ReLU& \\
 \hline 
 2&&2&&sigmoid&X \\
 \hline
 3&& 3 &X& tanh&\\
 \hline
 \end{tabular}
 \begin{tabular}{|cc|cc|cc|}
 \hline
 \textbf{Dropout Rate}&&\textbf{Number of First Layer Kernels}&&\textbf{First Layer Filter Size}& \\
 \hline 
 0.1 & & 4 & &1& \\
 \hline 
 0.3 &X& 8 &X &3& \\
 \hline
 0.5&& 16 & & 5& X\\
 \hline
 0.7&& 32 & & 7 & \\
 \hline
 0.9&&64 & & 9 &\\
 \hline
 \end{tabular}
\end{table*}

\subsection{Hyperparameter Tuning}

In order to produce a model with optimal accuracy, many different configurations of hyperparameters of the neural network were tested. These include basic hyperparameters such as which types of layers were used, the number of layers, and the order of the layers in the model. This constitutes the basic structure of our model. Other optimized attributes include features of each layer such as the number of kernels, filter sizes, and strides in the convolutional layers, and the dropout rate and output activation function. The results of these hyperparameter tuning tests are explained further in 3.2.

The first important hyperparameter testing was done to determine the amount of convolutional layers to be used, which was tested with the amount of max pooling per pooling layer, with a max pooling size of 1 not having an effect on the model. It was important to test how many convolutional layers were necessary, as too few layers would not be able to pick up on the complexities of the spectral data features. On the other hand, having too many convolutional layers would add an unnecessary computation to the model and even reduce the sensitivity of the model. For max pooling size, it was crucial because while a pooling layer helps reduce the amount of computation, too much of pooling would take out key information from the feature maps which would reduce the final sensitivity of the CNN in detecting key features.

Kernel shapes are the sizes of the filters that the convolutional layer uses. Kernel shapes for each convolutional layer were chosen by estimating the size of the features (spectral lines) we would be looking for \citep{Zhao2019}. Kernel count, or the number of filters applied, is the other key characteristic, which was determined in hyperparameter tuning tests. The number of kernels affects the ways that the layer can represent the features in the spectrum. If there are too many kernels, there would be a high possibility of overfitting. If there are too few kernels, the feature maps produced would not be enough to represent various aspects of the patterns of the spectral lines. Because the convolutional layer requires the movement of the filter along the data points, the stride is a parameter used to define the amount of movement each time, or the number of sections the layer will look at throughout the spectrum. Therefore, if the stride is too large, it is easy for the convolutional layer to skip some of the details. However, if the stride is too small, the computation time would be quite large.

The dropout rate is the fraction of data in the input layer that would be removed towards the end. If the dropout rate is too small, it would defeat the purpose of the layer, which is to prevent overfitting. However, if the dropout rate is too large, many of the important features in the data could be removed. The output activation is used as part of the final classification to determine what the eventual model score is. Different output activations will use different functions and obtain different classification results, so it is important to know which output activation achieves the best accuracy.

Extensive tuning of the aforementioned hyperparameters was conducted with grid search and cross-validation \citep{Dangeti2017} using the training and validation sets. Some of the hyperparameters tested and the resulting selections are shown in Table \ref{tab:hyperparameters}.

\begin{figure}
 \includegraphics[width=0.45\textwidth]{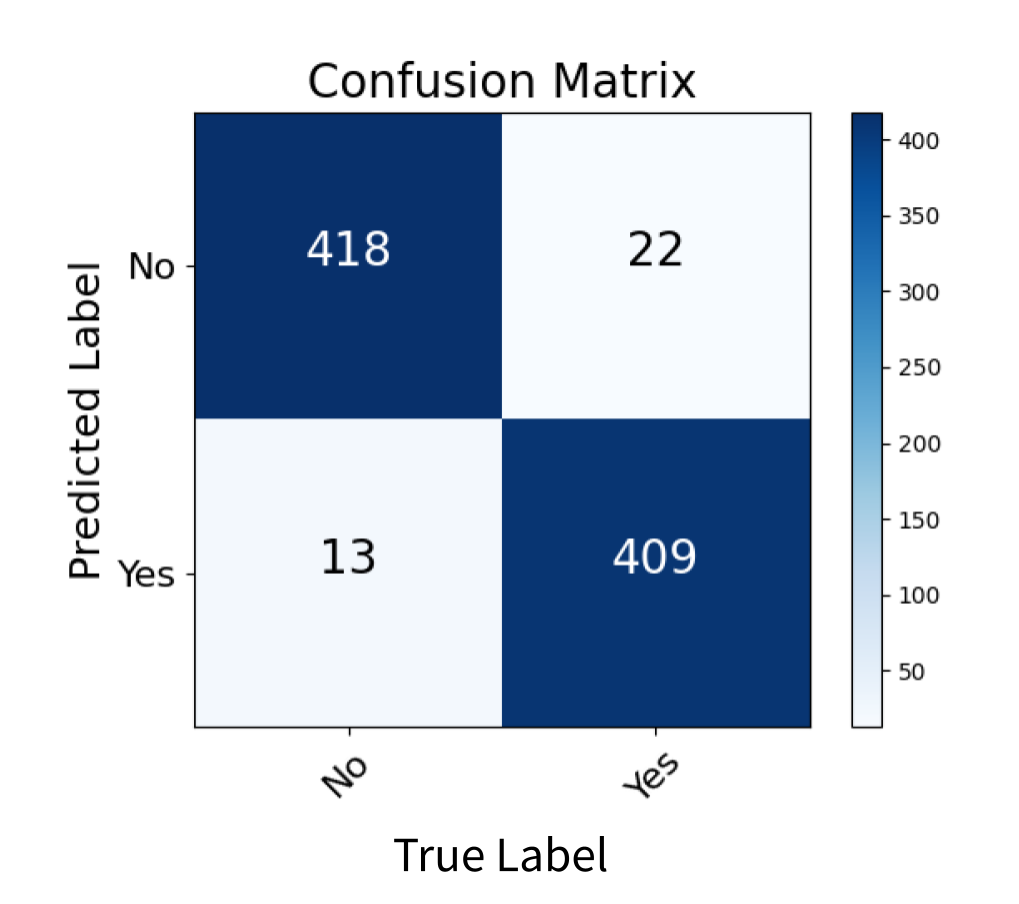}
 \caption{The confusion matrix of applying the model on the test set.}
 \label{fig:figconf}
\end{figure}

\begin{figure*}
 \includegraphics[width=1.0\textwidth]{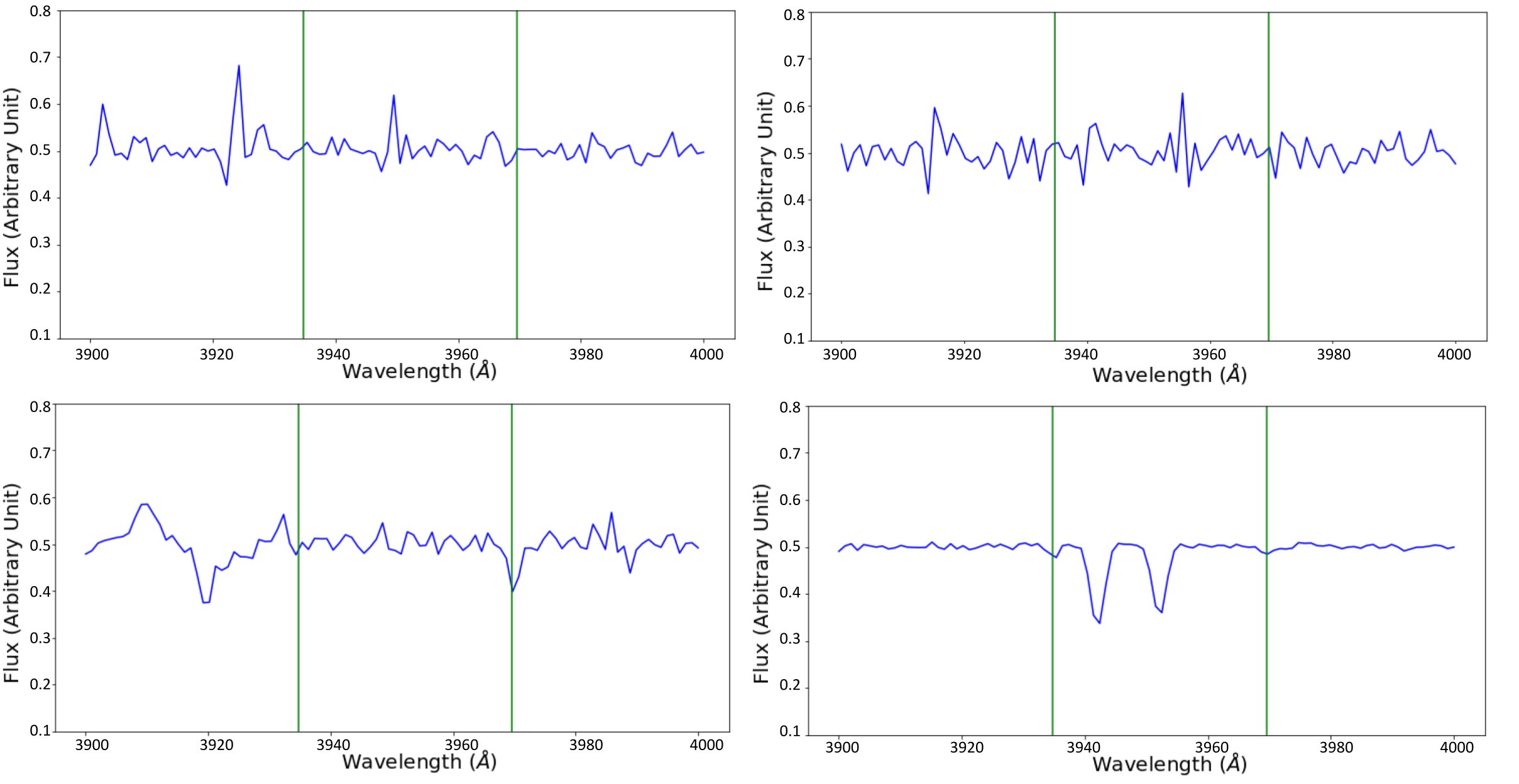}
 \caption{Examples of false negatives that are in the catalog of \citet{Sardane2014} plotted after both spectrum and noise normalization (after preprocessing). Based on our measurements, we believe that all but the Ca II absorption lines in the bottom right are actually real negatives. Top Left: Quasar J005355.15-000309.3 where the two absorption lines are not centered at 3934.78\r{A} and 3969.59\r{A}, respectively, making it possible that this is in fact not a Ca II absorber. Top Right: For quasar J012412.47-010049.7, both the \(\lambda\)3934 line and the \(\lambda\)3969 line are off-center in different directions. Furthermore, both lines appear to be narrow and weak which may be produced by random noises. Bottom Left: Because the \(\lambda\)3934 line especially is rather weak, quasar J101748.68+222659.2's absorption lines are not strong enough for it to be considered as having Ca II absorption lines. Bottom Right: Quasar J160335.78+453656.3 which is one of the- real absorption lines that the model missed. It is most likely due to the fact that the two absorption line in the middle leads the model to believe that the two lines are too weak, especially after noise normalization.}
 \label{fig:figsardane}
\end{figure*}

\subsection{Details about Our Model Design and Tuning}

Our model is built in Python with the open-source libraries Keras and TensorFlow. The input spectra have a 100-pixel length representing the flux between the wavelengths of 3900 \r{A} and 4000 \r{A}. It features five major layers (excluding normalization, activation, and dropout); three convolutional layers and two fully-connected layers (Figure \ref{fig:figarc}). We also decided to not include any max-pooling layers because due to our relatively low dimension and highly information-dense starting input, reducing this information further in this manner is more destructive than helpful. We included both an activation layer and a batch normalization layer and found that when placing the batch normalization layer after the activation function layer, the accuracy improved by about 1\%.

In the first convolutional layer, the 1-D kernel size is 5 pixels, which is approximately the average width of the spectral lines, plus a data point at each side. We also found that when the filter size was too high (9 pixels) or too low (3 pixels), the accuracy would decrease by around 2\%. Additionally, a filter count of 8 was chosen for the first layer to be able to best represent all of the intricacies of the spectrum including the two spectral lines, line locations and separations, line widths, depths, and profiles of the two spectral lines, and noise characteristics. The other 2 convolutional layers have a filter count of 16 and 32, respectively, to capture more details of the characteristics of both lines. Moreover, tests using a greater amount of kernels did not show much of a difference in model performance, which is not surprising as there is limited information to represent these two spectral lines and their variations. A stride (or a moving step) of 1 for each convolutional layer was chosen to look through all of the important spectral features within our data because the input size was relatively small so computation efficiency was not a problem. We also chose a 30\% dropout rate because we found that due to our small input size, it was necessary that the majority of the input still remain at the end. We used the rectified linear unit (ReLU) activation function after each of the convolutional layers and first fully connected layer (i.e. dense or densely connected layer), as it proved to be the most effective activation function between models due to its simplicity. However, for our output activation, we used the sigmoid function, which is the most often used output activation, because it converts the model output to a probability score between 0 and 1.

For training, the Adaptive Moment Estimation (Adam) \citep{Kingma2015} optimizer was chosen with a mini-batch size of 64. This mini-batch size, which is the number of training samples per iteration is usually chosen through hyperparameter tuning after the model is created. While a mini-batch size of 32 is usually the default, we found that we had a higher accuracy by about 1\% with a mini-batch size of 64. The Adam optimizer computes adaptive learning rates for each parameter, which effectively mitigates the fluctuation of the cost function during training, and keeps the training more stable. Test results and the Receiver Operating Characteristic (ROC) curve of the model are shown in Figure \ref{fig:figthreshold}. Since the \textit{true} scores are concentrated near 1.0 and \textit{false} scores are comparatively spread out, the true-false separation is set at 0.9, as indicated by the red marker in the ROC curve.

\section{Results}
Four metrics are used to evaluate the classification results of our model: accuracy, precision, recall rate, and F$_1$ score, which are defined below.

\textbf{Accuracy}: the percentage of correct predictions out of the total predictions, including both positive and negative samples i.e. Accuracy = (true positives + true negatives)/(true positives + true negatives + false positives + false negatives).

\textbf{Precision}: the fraction of samples correctly identified as positive out of all samples predicted as positives i.e. Precision = true positives/(true positives + false positives)

\textbf{Recall Rate}: the ratio between the number of correct predictions of a positive sample to the total number of positive samples, which indicates the sensitivity of the classifier i.e. Recall Rate = true positives/(true positives + false negatives)

\textbf{F$_1$ Score}: the harmonic mean of the precision and the recall rate, i.e. F$_1$ Score = 2*(Precision*Recall)/(Precision + Recall)

The F$_1$ score favors classifiers that have a similar precision and recall rate, thus is a better measure to use when seeking a balance between precision and recall.

Our experimental results on the test set are summarized in Table \ref{tab:tabresults} and the confusion matrix is shown in Figure \ref{fig:figconf}. Overall, all the evaluation metrics of our model are above 94.9\% with the precision a little higher than the recall. We were also able to discover a large number of new quasar Ca II absorbers with the model. More details on the results are described in the following subsections.

\begin{table}
 \centering
 \caption{The classification metrics of our model on the test set with different prediction thresholds. A threshold of 0.90 was chosen which gave the highest overall accuracy.}
 \begin{tabular}{c|c|c|c|c}
  \hline
   Threshold & 0.90 & 0.80 & 0.65 & 0.30 \\
   \hline
  Accuracy & 95.9\% & 95.7\% & 94.9\% & 92.0\% \\
  Precision & 96.9\% & 95.8\% & 93.5\% & 88.0\% \\
  Recall & 94.9\% & 95.6\% & 96.5\% & 97.2\% \\
  F$_1$ Score & 95.9\% & 95.7\% & 95.0\% & 92.4\% \\
   \hline
 \end{tabular}
 \label{tab:tabresults}
\end{table}

\begin{figure*}
 \includegraphics[width=\textwidth]{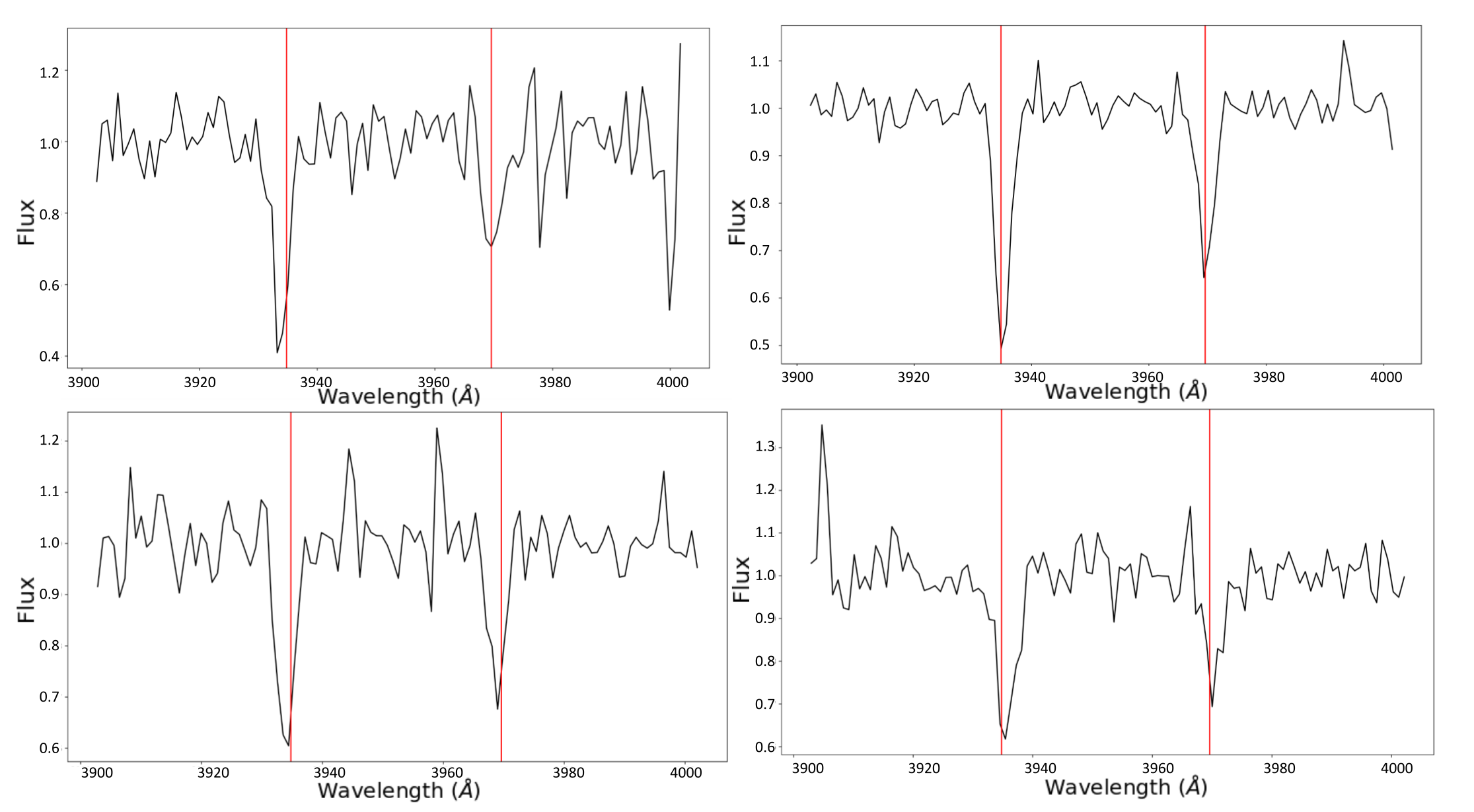}
 \caption{Examples of new quasar spectra with Ca II absorption lines that were discovered by our neural network. The examples on the left are from SDSS’s DR12 (Top Left: J100806.18+234942.1, Bottom Left: J120301.01+063441.5) and the examples on the right are from SDSS’s DR7 (Top Right: J122016.87+112628.1 Bottom Right: J065412.58+283007.2).}
 \label{fig:fignewabs}
\end{figure*}
\begin{figure*}
 \includegraphics[width=\textwidth]{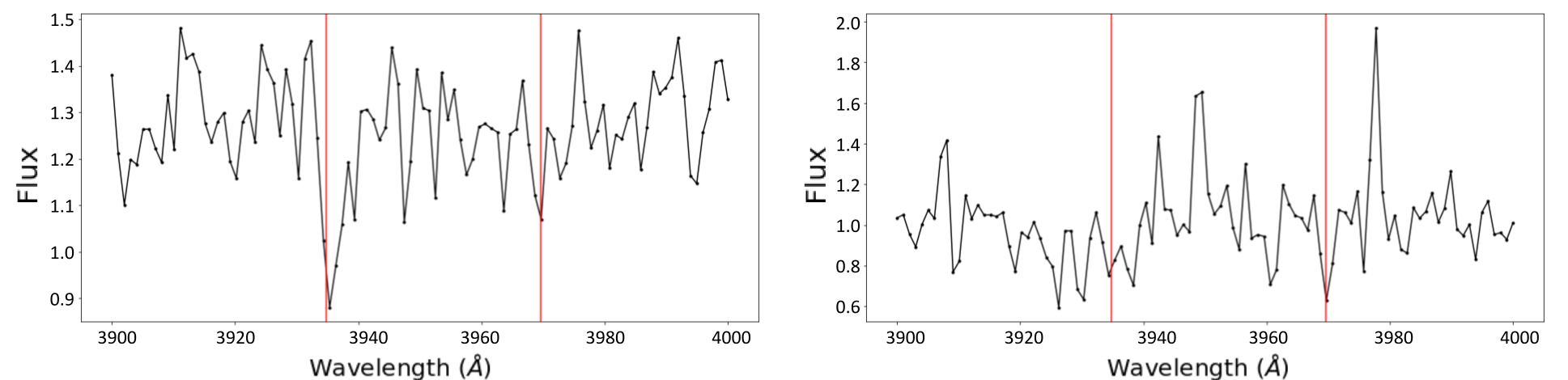}
 \caption{Examples of quasar spectra that were identified by the model but then manually discarded. The one on the left has a strong \(\lambda\)3934 absorption line but an insignificant \(\lambda\)3969 absorption line. Meanwhile, the one on the right has plenty of noise, making both lines insignificant.}
 \label{fig:figfalsepos}
\end{figure*}
\begin{table}
  \centering
  \caption{The 13 quasars in \citet{Sardane2014}'s that we believe may be false detections.}
  \begin{tabular}{c}
    \hline
    SDSS name \\ \hline
    J005355.15-000309.3 \\ 
    J012412.47-010049.7 \\
    J094613.97+133441.2 \\
    J101748.68+222659.2 \\
    J124347.60+374512.5 \\
    J133526.01-010028.1 \\
    J134246.25-003543.7 \\
    J142119.39+313219.6 \\
    J152941.57+254815.9 \\
    J155453.30+245622.5 \\
    J103451.42+233435.4 \\
    J085223.93+565725.7 \\
    J165118.61+400124.8 \\\hline
  \end{tabular}
  \label{tab:my_label}
\end{table}
\subsection{Results on the Existing Catalog}
Out of the 431 spectra in our positive test set from \citet{Sardane2014}'s catalog, our model was able to accurately identify 409 quasar spectra with Ca II absorbers, giving a 94.9\% recall rate. In our model, we categorized absorption lines as real Ca II absorption lines or candidate ones based on the SNR of the two absorption lines at \(\lambda\)3934 and \(\lambda\)3969. The real Ca II absorption lines are defined as having an absorption line at \(\lambda\)3934 with a SNR > 3 and an absorption line at \(\lambda\)3969 with a SNR > 2.5, and candidate Ca II absorption lines as having an absorption line at \(\lambda\)3934 with a SNR of 2.5-3 and an absorption line at \(\lambda\)3969 with a SNR of 2.0-2.5. 

We conducted analysis on the remaining 22 quasar spectra which were not identified by our model as Ca II absorber detections by measuring the EWs and EW errors of potential Ca II absorption lines and found that only 5 quasar spectrum has real Ca II absorption lines and 4 quasar spectra have candidate Ca II absorption lines missed by our model. Our model most likely missed these 9 quasar Ca II absorbers due to narrow lines or other factors such as the strong absorption lines nearby in the bottom right spectrum of Figure \ref{fig:figsardane}. The remaining 13 quasar spectra have no absorption lines meeting our definitions of Ca II absorbers or candidate absorbers (Table \ref{tab:my_label}). Their absorption lines are either not centered at 3934.78 \r{A} and 3969.59 \r{A} such as in the top left spectrum of Figure \ref{fig:figsardane}; or too narrow and weak, causing them to not be identified by the program, such as in the bottom left spectrum of Figure \ref{fig:figsardane}; or where both issues occur, such as in the top right spectrum of Figure \ref{fig:figsardane}. It is likely that these absorption lines were simply false detections.

\begin{table*}
\centering
\caption{Some examples from our new catalog of 399 quasars with Ca II absorption lines, \textbf{with the entire catalog listed online as supplementary information.}}
\begin{tabular}{|c|c|c|c|c|c|c|c|c|}
\hline
\textbf{SDSS name} &\textbf{RA (degrees)} & \textbf{Dec (degrees)} & \textbf{zabs} & \textbf{zqso} & \textbf{EW[3934]} (\AA) & \textbf{Err[3934]} (\AA)& \textbf{EW[3969]} (\AA) & \textbf{Err[3969]} (\AA) \\ \hline
J105220.99+183636.7 & 163.087 & 18.610 & 0.959  & 1.091   & 0.8    & 0.29     & 0.51    & 0.17     \\ \hline
J094636.86+323949.5& 146.654 & 32.664 & 0.798  & 1.308  & 0.61    & 0.06     & 0.36    & 0.05     \\ \hline
J132323.78-002155.2 & 200.849 & -0.365 & 0.716  & 1.392  & 1.04   & 0.10     & 0.48    & 0.08     \\ \hline
J081739.19+453228.3 & 124.413 & 45.541 & 0.749  & 1.510  & 0.95    & 0.10     & 0.58    & 0.09     \\ \hline
J090122.67+204446.5 & 135.344 & 20.746 & 1.019  & 2.103  & 0.48    & 0.06     & 0.27    & 0.04     \\ \hline
J065412.58+283007.2 & 103.552 & 28.502 & 0.630  & 1.634  & 1.52    & 0.18     & 0.86    & 0.14     \\ \hline
J075123.60+084248.8 & 117.848 & 8.714 & 0.543  & 1.550 & 0.84    & 0.09     & 0.41    & 0.09      \\ \hline
J072400.03+320226.6 & 111.000 & 32.041 & 0.572  & 1.159  & 1.50    & 0.19     & 0.81    & 0.14     \\ \hline
J094145.03+303503.6 & 145.438 & 30.584 & 0.937  & 1.226  & 0.93    & 0.12     & 0.64    & 0.08     \\ \hline
J122016.87+112628.1 & 185.070 & 11.441 & 0.731  & 1.896  & 1.59    & 0.10     & 1.10    & 0.13     \\ \hline
\end{tabular}
\label{tab:CaIIcatlog}
\end{table*}
On the precision rate side, our model predicted 422 positive predictions in total, and out of them, was able to correctly classify 409 out of the 422 positive predictions, achieving a 96.9\% precision. Within the 13 false positives, many had a large amount of noise or absorption lines very close to \(\lambda\)3934 and \(\lambda\)3969 which the model would mark as real absorption lines. Weak lines were also often present in these false detections.

\begin{figure*}
 \includegraphics[width=0.5\textwidth]{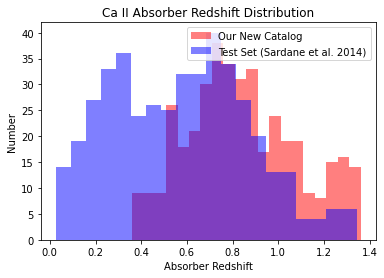}
 \caption{The comparison between the absorber's redshift distributions of our test set from \citet{Sardane2014}, marked in blue color, and our new catalog marked in red color. Our new catalog has a more confined redshift distribution (i.e. \(z_{abs}\)>0.36) due to the dependence on Mg II absorption lines. Overall, there are significantly more absorbers with mid to higher redshift and a greater average redshift in our new catalog than what is found in \citet{Sardane2014}.}
 \label{fig:figredcomp}
\end{figure*}

\subsection{Results on DR7 and DR12}
Because our preprocessing method requires redshifts of quasar Mg II absorbers to locate potential Ca II absorption lines, only quasar spectra with Mg II absorption lines were searched for Ca II absorption lines, resulting in Ca II absorber's redshifts in the range of 0.36 < \(z_{abs}\) < 1.4. In total, there are 35,752 quasar spectra with Mg II absorption lines in SDSS’s DR7 identified by \citet{2013ApJ...770..130Z} and 41,895 quasar spectra with Mg II absorption lines in DR12 found by \citet{Zhao2019}. We then limited to those quasar spectra that have Mg II's redshifts within the range that Ca II absorption lines can be found, i.e., with 0.36 < \(z_{abs}\) <1.4, which left a total of 24,827 quasars from DR7 and 18,418 quasars from DR12. The neural network model ran through these two sets separately, identifying 1,267 spectra with potential Ca II absorption lines in DR7 and 694 spectra in DR12. 

The EW of the \(\lambda\)3934 line and the \(\lambda\)3969 line of every candidate was measured by normalizing each of the spectra from 3900 \r{A} to 4000 \r{A} and then fitting each of the lines individually through a Gaussian fit. Real absorption lines and candidates are those meeting the SNR requirements described in section 4.1. A total of 399 new Ca II absorbers were identified (137 in DR7 and 262 in DR12). Examples of these absorption lines are shown in Figure \ref{fig:fignewabs}. A sample of absorbers and their properties are listed in Table \ref{tab:CaIIcatlog}.

\begin{figure*}
 \includegraphics[width=\textwidth]{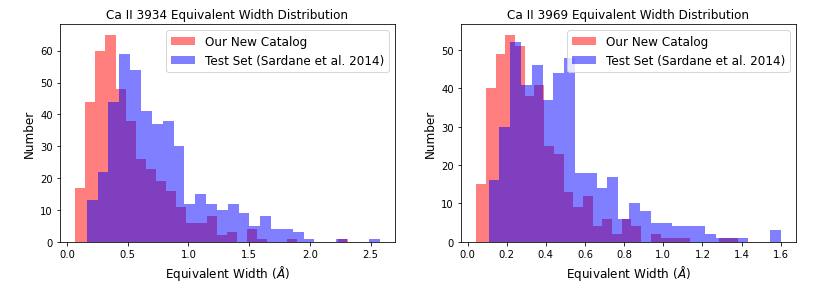}
 \caption{The comparison between the EW distributions of the \citet{Sardane2014} set (our test set), marked in blue color, and our new catalog, marked in red color. Left: The comparison of the EW distributions for the \(\lambda\)3934 line. Right: The comparison of EW distributions for the \(\lambda\)3969 line. Our catalog has a broader range for both absorption lines and is more skewed to cover lower EW values, indicating that our approach is able to discover weaker absorption lines.
}
 \label{fig:figewcomp}
\end{figure*}
\begin{table*}
 \centering
 \caption{The comparison of the distributions of the absorption redshift, equivalent width for the \(\lambda\)3934 line, and equivalent width for the \(\lambda\)3969 line of \citep{Sardane2014}'s catalog with our catalog.}
 \begin{tabular}{c | c | c | c | c}
  \hline
  Measurement & Sardane's Catalog Distribution & Mean $\pm$ Standard Deviation & Our Catalog Distribution & Mean $\pm$ Standard Deviation \\
  \hline
  Absorption Redshift & 0.03 < \(z_{abs}\) < 1.34 & 0.58 $\pm$ 0.30 & 0.36 < \(z_{abs}\) < 1.36 &0.84 $\pm$ 0.24\\
  Equivalent Width \(\lambda\)3934  (\AA)& 0.16 < EW < 2.57 & 0.76 $\pm$ 0.39 &0.07 < EW < 2.31 & 0.51 $\pm$ 0.31 \\
  Equivalent Width \(\lambda\)3969  (\AA) & 0.11 < EW < 1.60 & 0.48 $\pm$ 0.26 & 0.04 < EW < 1.38 & 0.33 $\pm$ 0.20\\
   \hline
 \end{tabular}
 \label{tab:tabcomparison}
\end{table*}
\begin{figure}
 \includegraphics[width=0.48\textwidth]{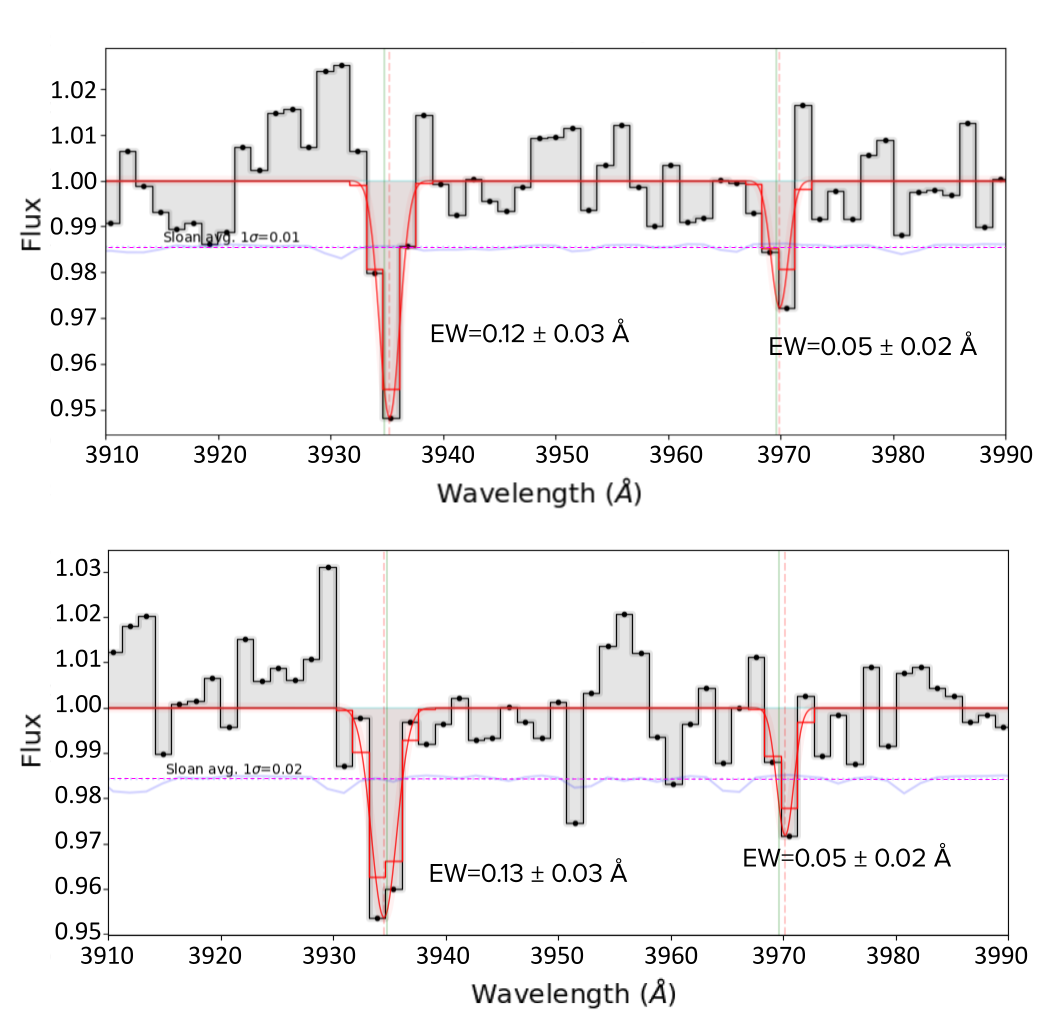}
 \caption{Examples of equivalent width fittings of weak absorption lines that our model was able to identify. The blue solid line shows the noise from the Sloan Digital Sky Survey catalog, and the pink dotted line is the average of the noise from the 3900 \r{A} - 4000 Å. The green lines show the locations of 3934.78 Å and 3969.59 Å, with the orange lines showing the centers of the equivalent width fittings for each line. Top: J142106.86+533745.4 which has a pair of Ca II absorption lines with a \(\lambda\)3934's EW of 0.12 \AA, and a \(\lambda\)3969's EW of 0.05 \AA. Bottom: J142106.86+533745.4 which has a pair of Ca II absorption lines with a \(\lambda\)3934's EW of 0.13 \AA\ and a \(\lambda\)3969's EW of 0.05 \AA.}
 \label{fig:weakAbsorbers}
\end{figure}

Some of the samples identified by the neural network but then discarded in the manual process have only one strong line with the other line either being too weak or nonexistent. Other samples have some sort of lines at \(\lambda\)3934 and \(\lambda\)3969 that could have been caused by external noise, with a large amount of error. It is possible that the artificial training set may have contained too big of a portion of weak absorbers to accommodate the EW distribution of the real data set, causing many false positives. Examples of these false absorption lines are shown in Figure \ref{fig:figfalsepos}.

\subsection{Comparison of the New Catalog with the Past Catalogs}
Our new catalog has a total of 399 quasar Ca II absorbers. A summary of measurement comparisons are shown in Table \ref{tab:tabcomparison}, redshift distribution and EW distributions are shown in Figures \ref{fig:figredcomp} and \ref{fig:figewcomp}. Our redshift distribution, compared to the redshift distribution of \citet{Sardane2014}, misses quasar spectra that have Ca II absorbers with a redshift in the range of 0.03 < \(z_{abs}\) < 0.36, due to the reliance on Mg II absorption lines to find Ca II absorption lines in the rest frame (Figure \ref{fig:figredcomp}). This situation could be changed in our future work by doing a direct survey of Ca II absorption lines in quasar spectra without using the Mg II absorber's redshift as a signpost. Nevertheless, we were also able to find a greater amount of absorption lines at higher redshifts in quasar spectra. This is most likely due to how our preprocessing method emphasizes normalizing the noise around Ca II absorption lines which allowed our model to be able to detect absorption lines even at higher redshifts where there is a greater amount of sky emissions at >8500 \r{A} in the observer wavelength. 

\begin{figure*}
 \includegraphics[width=\textwidth]{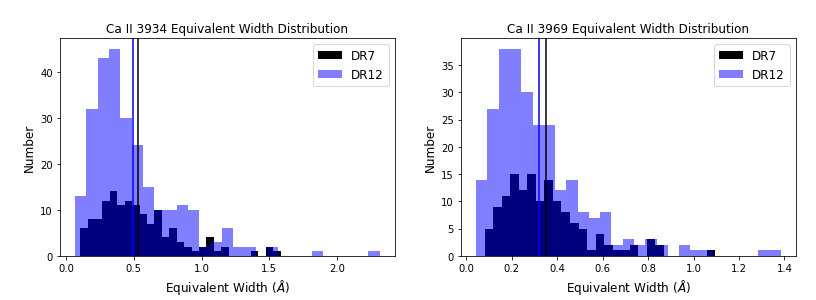}
 \caption{The comparison between the EW distributions of the \(\lambda\)3934 and \(\lambda\)3969 lines for quasar Ca II absorbers in our catalog that were discovered in DR7 and those that were discovered in DR12. Left: The EW distribution for \(\lambda\)3934, with DR7 in black color and DR12 in blue color. The means of the EWs for both the DR7 and DR12 are also shown in their respective colors, with 0.52 \AA\ for DR7 and 0.49 \AA\ for DR12. Right: The EW distribution for \(\lambda\)3969, with DR7 in black color and DR12 in blue color. The means of the EWs for both the DR7 and DR12 are also shown in their respective colors, with 0.36 \AA\ for DR7 and 0.32 \AA\ for DR12.}
 \label{fig:DR7DR12}
\end{figure*}

On the other hand, EW distributions for both the \(\lambda\)3934 and the \(\lambda\)3969 lines cover much bigger ranges and are overall more skewed to weaker absorbers than those reported in \citet{Sardane2014}. This means that our model is able to successfully discover absorption lines with lower EW values, therefore allowing it to find weaker absorbers. We are able to find weak absorption lines because our model was trained on absorption lines that have a great range in EWs, including many weak ones. These weak absorption lines are also likely harder to be examined. Examples of weak absorption lines are shown in Figure \ref{fig:weakAbsorbers}. Since most strong absorption lines in both DR7 and some in DR12 (as DR9 used in \citet{Sardane2014} is part of DR12) have been identified by \citet{Sardane2014}), the overall samples in our catalog appear to show weaker absorption lines. When comparing our EW distributions for DR7 and DR12 for both the \(\lambda\)3934 and \(\lambda\)3969 line, DR12 appears to have an overall lower EW distribution than DR7, which is shown in Figure \ref{fig:DR7DR12}. This trend remained even after we included all confirmed samples from \citet{Sardane2014}, who discovered most of the strong DR7 absorption lines. Because we have a greater amount of quasars in DR12, it is logical why our catalog may have weaker absorption lines. This also demonstrates the power of our model to be able to find both strong and weak absorption lines in catalogs that were already searched.

\section{Conclusion and Discussion}
Overall, in this work, we were able to successfully develop a neural network model that can detect Ca II absorption lines with an accuracy of 95.9\%. Our proposed approach could accurately discover Ca II absorption lines, including almost all of the absorption lines identified in previous catalogs, as well as new absorbers. Furthermore, its detection speed is much faster than traditional methods. When looking through traditional methods as we did in \citet{Fang2022}, we found that it takes many days to a couple of weeks to look through about 10,000 spectra. However, with our method, the neural network can analyze about 10,000 spectra in only 0.7 seconds. This makes our method tens of thousands of times faster than traditional methods.

In our approach, a large amount of artificial quasar Ca II absorption spectra used for neural network training has resolved the shortage issue of real Ca II absorber samples and helped improve the accuracy of the neural network.
Our technique of moving the absorption lines to the rest frame and limiting the neural network search window to a relatively small fixed wavelength window appears to have further simplified our neural network design and increased the accuracy and speed of the model. Our data preprocessing involving the noise normalization largely improves the accuracy and effectiveness in searching Ca II absorption lines at wavelengths where there is a great amount of sky emission noise, which used to be an especially challenging situation. These techniques have not been reported in any other neural network model for discovering absorption lines to our best knowledge. This is also the first time that deep learning has been applied to discovering Ca II absorption lines.

Our model was tested on SDSS’s DR7 and DR12, both of which were already partially searched by \citet{Sardane2014} who inspected DR7 and DR9 (which overlaps with DR12). With our developed neural network model, we have been able to confirm a total of 409 spectra with Ca II absorption lines that were found in the past. We have also been able to find a total of 399 new Ca II absorption lines in DR7 and DR12, which is over 100 greater than what we were able to find using a traditional method in \citet{Fang2022}. Furthermore, 137 new Ca II absorption lines were found in DR7 which had already been previously searched by \citet{Sardane2014}. Of these 137, it is unsurprising that the majority are weak (\(\lambda\)3934 EW <0.7 \AA), with about 107 absorption lines being weak, and 30 being strong. Most of these new absorption lines have lower EWs, indicating the capability of our model on detecting weaker Ca II absorption lines.

When looking into the 22 absorption lines reported in \citet{Sardane2014} but missed by our model, we found that 13 of them are false signals, often having misplaced absorption lines or absorption lines that were not significant enough. This demonstrates the ability of our model to discover false absorption lines from existing work. Using our model, we have been able to confirm and discover a total of 808 Ca II absorption lines altogether among quasar spectra from DR7 and DR12 (399 new Ca II absorption lines and 409 previously found absorption lines), greatly adding to the currently available set of found Ca II absorption lines that can be then analyzed to understand dust absorbers and chemical abundances of recent galaxies.

As stated previously, prior studies have shown that all quasars with Mg II absorption lines have Ca II absorption lines for permitting SNR in the Mg II region, and with a \(z_{abs}\) > 0.4. Previous studies such as Sardane et al. (2014), have had a minimum EW of 0.08 Å for its corresponding Mg II \(\lambda\)2796 line. In fact, the minimal EW of the Mg II \(\lambda\)2796 line for all previously detected Ca II absorbers has been much larger than 0.08 Å.  This indicates that it is extremely possible that Ca II absorbers can only be detected with Mg II absorbers with a \(\lambda\)2796 EW of at least 0.08 Å. In our study, since we have looked at quasars with Mg II absorption lines with a minimum EW of 0.07 Å for its \(\lambda\)2796 line, it is highly likely that  we have searched through all quasars that Ca II absorption lines could be in with a \(z_{abs}\) > 0.4, giving us reasons to believe that we have found almost all Ca II absorption lines with \(z_{abs}\) > 0.4 in DR7 and DR12.

In comparison with \citet{Zhao2019}, which had successfully identified over 40,000 Mg II absorption lines in SDSS's DR12 quasar spectra using a neural network for the first time, we were able to use a neural network to detect Ca II absorption lines for the first time. These Ca II absorption lines are much weaker and rarer than Mg II absorption lines, making them more difficult to be identified in quasar spectra. Our newly developed preprocessing techniques has helped identify these Ca II absorbers, especially those weak ones, with high accuracy and search speed. 

Nevertheless, our approach with the neural network does have room for improvement. There were many false absorption lines detected by the model from both DR7 and DR12, most likely due to its misclassification of noise as absorption lines. It also tends to miss absorption lines that appear noisier and narrower. Training sets could also only include absorption lines with a good \(\lambda\)3934 to \(\lambda\)3969 line ratio of around 2:1 to possibly weed out more of the false positives. Moreover, our processing approach currently requires the use of each quasar’s absorber redshift value before searching through it, limiting the possible spectra it is able to search, as it is necessary to find the absorber redshift from the Mg II absorption lines in the quasar. In the future, we plan to explore a different preprocessing technique to be able to find Ca II absorption lines in the observer frame, so that the absorber’s redshift is not required. For example, it is possible to use a combination of a sliding window and a neural network to inspect local regions of the spectrum where Ca II absorption lines may potentially exist and solve for the absorber’s redshift this way. Alternatively, the entirety of the spectrum may be used such as with \citet{Zhao2019} as input to the neural network but then would need to deal with the extra processing and noises involved. We also plan to continue using our neural network to discover more Ca II absorption lines in DR14 and later data releases from SDSS that have not been processed so as to continuously expand the current catalogs of Ca II absorption lines.

\section*{Acknowledgements}
Funding for the SDSS and SDSS-II has been provided by the Alfred P. Sloan Foundation, the Participating Institutions, the National Science Foundation, the U.S. Department of Energy, the National Aeronautics and Space Administration, the Japanese Monbukagakusho, the Max Planck Society, and the Higher Education Funding Council for England. The SDSS Web Site is http://www.sdss.org/.

The SDSS is managed by the Astrophysical Research Consortium for the Participating Institutions. The Participating Institutions are the American Museum of Natural History, Astrophysical Institute Potsdam, University of Basel, University of Cambridge, Case Western Reserve University, University of Chicago, Drexel University, Fermilab, the Institute for Advanced Study, the Japan Participation Group, Johns Hopkins University, the Joint Institute for Nuclear Astrophysics, the Kavli Institute for Particle Astrophysics and Cosmology, the Korean Scientist Group, the Chinese Academy of Sciences (LAMOST), Los Alamos National Laboratory, the Max-Planck-Institute for Astronomy (MPIA), the Max-Planck-Institute for Astrophysics (MPA), New Mexico State University, Ohio State University, University of Pittsburgh, University of Portsmouth, Princeton University, the United States Naval Observatory, and the University of Washington.

Funding for SDSS-III has been provided by the Alfred P. Sloan Foundation, the Participating Institutions, the National Science Foundation, and the U.S. Department of Energy Office of Science. The SDSS-III web site is http://www.sdss3.org/.

SDSS-III is managed by the Astrophysical Research Consortium for the Participating Institutions of the SDSS-III Collaboration including the University of Arizona, the Brazilian Participation Group, Brookhaven National Laboratory, Carnegie Mellon University, University of Florida, the French Participation Group, the German Participation Group, Harvard University, the Instituto de Astrofisica de Canarias, the Michigan State/Notre Dame/JINA Participation Group, Johns Hopkins University, Lawrence Berkeley National Laboratory, Max Planck Institute for Astrophysics, Max Planck Institute for Extraterrestrial Physics, New Mexico State University, New York University, Ohio State University, Pennsylvania State University, University of Portsmouth, Princeton University, the Spanish Participation Group, University of Tokyo, University of Utah, Vanderbilt University, University of Virginia, University of Washington, and Yale University.

The authors would like to thank the anonymous referee who has provided valuable suggestions which help improve the quality of this paper.

\section*{Data Availability Statement}

The data underlying this article were accessed from the Sloan Digital Sky Survey (https://classic.sdss.org/). The data generated from this research are available in the article and in its online supplementary material.



\bibliographystyle{mnras}
\bibliography{example} 



\bsp	
\label{lastpage}
\end{document}